# Brownian Motion of a Classical Particle in Quantum Environment

Roumen Tsekov

Department of Physical Chemistry, University of Sofia, 1164 Sofia, Bulgaria

The Klein-Kramers equation, governing the Brownian motion of a classical particle in quantum environment under the action of an arbitrary external potential, is derived. Quantum temperature and friction operators are introduced and at large friction the corresponding Smoluchowski equation is obtained. Introducing the Bohm quantum potential, this Smoluchowski equation is extended to describe the Brownian motion of a quantum particle in quantum environment.

The Brownian motion is a central phenomenon of statistical mechanics. Robert Brown discovered it originally on colloidal particles but nowadays it is evident that any particle in a thermodynamic system undergoes Brownian movement. The onset of quantum mechanics definitely affected the Brownian motion and, in general, the quantum Brownian motion describes a quantum particle surrounded by quantum environment [1]. However, due to mathematical difficulties, it is impossible to solve explicitly the complete quantum dynamics of the entire thermodynamic system. For this reason, two semi-classical models are proposed as well. The first one considers motion of a quantum particle in classical environment, which is relevant, for instance, to the important motion of an electron in metal. The well-known Caldeira-Leggett model [2] is valid only at high temperature and certainly belongs to this class. The second model describes motion of a classical particle in quantum environment. It is rigorously derived for Brownian oscillators, since in this case the dynamic force is linear [1]. Thus, one can employ the convenient Fourier transform on time. The scope of the present article is to develop an approach to the Brownian motion of a classical particle in quantum environment under the action of arbitrary external potentials. The corresponding Klein-Kramers equation is derived by an original temperature operator, which acts via time derivatives, similar to the quantum energy operator $\hat{E} \equiv i\hbar\partial_t$.

The stochastic Langevin equation describes most generally the Brownian motion of a classical particle with mass $m$ and coordinate $R(t)$ in the field on an external potential $U(R)$

$$m\ddot{R} + m\gamma\dot{R} = -\partial_R U + F \tag{1}$$

The fluctuation-dissipation theorem [1] relates the random and dissipative forces in such a way that the Langevin force spectral density $S_{FF} = m\gamma I \hbar\omega \coth(\beta\hbar\omega/2)$ is proportional to the friction tensor $m\gamma I$, where $I$ is the unit one and $\beta \equiv 1/k_B T$ is the reciprocal temperature. Since $F$ is a zero centered Gaussian stochastic process, its autocorrelation function $C_{FF}$ determines statistically all properties of the Langevin force. Inverting this Fourier image yields the Langevin force autocorrelation function

$$C_{FF} \equiv <F(t)F(0)> = 2m\gamma I k_B \hat{T} \delta(t) \tag{2}$$

where $\delta(t)$ is the Dirac delta-function. Equation (2) is valid for a general environment and in the classical limit $\hbar \to 0$ it simplifies to the well-known expression $C_{FF} = 2m\gamma I k_B T \delta(t)$ for a white noise. The novel temperature operator

$$\hat{T}/T \equiv (\beta\hat{E}/2)\coth(\beta\hat{E}/2) = \sum_{n=0}^{\infty} B_{2n}(\beta\hat{E})^{2n}/2n! = \cot(\beta\hbar\partial_t/2)(\beta\hbar\partial_t/2) \tag{3}$$

is defined via series expansion on $\hat{E} \equiv i\hbar\partial_t$, where $B_{2n}$ are even Bernoulli numbers. It is introduced for compactness and $\hat{T}$ should not be considered as an operator of quantum mechanics. One can check its correctness by applying Fourier transformation to Eq. (2). The resultant image $S_{FF} = 2m\gamma I k_B T \cot(\beta\hbar i\omega/2)(\beta\hbar i\omega/2)$ coincides with the fluctuation-dissipation theorem expression due to the properties of the trigonometric function. Taking the first two terms of Eq. (3), which are leading ones at high temperature, yields a semi-classical temperature operator $T + \beta\hat{E}^2/12k_B$. Note that there is no asymptotic form of the temperature operator (3) at zero temperature. In the traditional Fourier analysis of linear Eq. (1) it is usually accepted that the Langevin force spectral density equals to $S_{FF} = m\gamma I \hbar\omega$ at zero temperature, which is, however, true only if $\omega \neq 0$. On the other hand, the diffusion coefficient follows from the Green-Kubo relation at $\omega = 0$, where $S_{FF} = 2m\gamma I k_B T$ is completely classical. Fortunately, this uncertainty at zero temperature is only hypothetical, since the Third Law of thermodynamics states that the absolute zero is impossible to reach, i.e. $T > 0$.

Solving Eq. (1) is not possible in the common case. For this reason a Klein-Kramers equation is desired, which governs the evolution of the phase-space probability density $f(p,r,t)$ of the classical Brownian particle. Differentiating $f \equiv <\delta(p - m\dot{R})\delta(r - R)>$ on time and expressing the Brownian particle acceleration from Eq. (1) yields

$$\partial_t f + p \cdot \partial_r f / m - \partial_r U \cdot \partial_p f = \partial_p \cdot (\gamma p f) - \partial_p \cdot < F\delta(p - m\dot{R})\delta(r - R) > \qquad (4)$$

Since the Langevin force is Gaussian, one can apply the Furutsu-Novikov-Donsker formula [3-5] to express the last unspecified statistical moment in Eq. (4)

$$< F\delta(p - m\dot{R})\delta(r - R) > = -\int C_{FF}(t-s) \cdot \partial_p f(p,r,s) ds = -m\gamma k_B \hat{T} \partial_p f \qquad (5)$$

Substituting now Eq. (5) in Eq. (4) results in the desired Klein-Kramers equation

$$\partial_t f + p \cdot \partial_r f / m - \partial_r U \cdot \partial_p f = \gamma \partial_p \cdot (p f + m k_B \hat{T} \partial_p f) \qquad (6)$$

In the classical limit Eq. (6) reduces to the well-known classical Klein-Kramers equation, since $\hat{T}$ tends to temperature $T$ at $\hbar \to 0$.

In general, one could expect also a frequency-dependent friction coefficient, which will result in a time-dependent friction operator $\hat{\gamma}$ [6]. For instance, in the case of a Brownian particle, dissipating energy via normal friction and electromagnetic radiation [7], the friction operator acquires the form $\hat{\gamma} = \gamma - \tau \partial_t^2$, where $\tau = e^2 / 6\pi\varepsilon_0 m c^3$ is a specific time constant. Accordingly, one can generalize further Eq. (6) to obtain

$$\partial_t f + p \cdot \partial_r f / m - \partial_r U \cdot \partial_p f = \hat{\gamma} \partial_p \cdot (p f + m k_B \hat{T} \partial_p f) \qquad (7)$$

Since the temperature operator $\hat{T}$ acts solely on time, the equilibrium solution of Eq. (7) is always the classical Maxwell-Boltzmann distribution

$$f_{eq} \simeq \exp(-\beta p^2 / 2m)\exp(-\beta U) \tag{8}$$

Therefore, the quantum environment affects the relaxation dynamics of the classical Brownian particle but do not change its equilibrium properties. This is evident from the fact that Eq. (8) follows from the Boltzmann statistics as well, which is general for any kind of thermostat.

If the friction is strong, the Brownian motion is described well by the Smoluchowski equation, which can be derived from the Klein-Kramers one. Integrating Eq. (6) on the particle momentum yields the continuity equation

$$\partial_t \rho = -\partial_r \cdot (\rho V) \tag{9}$$

where $\rho \equiv \int f dp$ and $\rho V \equiv \int p f dp / m$ are the probability density and flow in the configurational space, respectively. Multiplying now Eq. (6) by the particle momentum and integrating along $p$ again yields after rearrangements the force balance

$$m\rho(\partial_t V + V \cdot \partial_r V) + m\hat{\gamma}(\rho V) = -\rho \partial_r U - \partial_r \cdot \Pi \tag{10}$$

where the osmotic pressure tensor is defined via $\Pi \equiv \int p \otimes (p/m - V) f dp$. In the case of strong friction, the first inertial term can be neglected and combining the remaining Eq. (10) with Eq. (9) results in

$$m\hat{\gamma}\partial_t \rho = \partial_r \cdot (\rho \partial_r U + \partial_r \cdot \Pi) \tag{11}$$

The pressure tensor can be calculated from Eq. (6) as well and in the considered case of strong friction it acquires the ideal gas form $\Pi = k_B \hat{T} \rho I$. Substituting this expression in Eq. (11) yields the desired Smoluchowski equation

$$m\hat{\gamma}\partial_t \rho = \partial_r \cdot (\rho \partial_r U + k_B \hat{T} \partial_r \rho) \tag{12}$$

As is expected, the equilibrium solution of this equation is the classical Boltzmann distribution. Note also that the quantum environment does not affect the mean kinetic energy of the classical particle, since $\int \mathrm{tr}(\Pi/2) dr = k_B \hat{T} 3/2 = 3k_B T/2$. This is not surprising, because the equilibrium occurs at infinite time $t \to \infty$, which corresponds to $\omega \to 0$ in the Fourier space, where the Langevin force spectral density tends to the classical expression. In the case of a free Brownian particle ($U \equiv 0$), Eq. (12) reduces to a generalized diffusion equation

$$m\hat{\gamma}\partial_t \rho = k_B \hat{T} \partial_r^2 \rho \tag{13}$$

Considering now the semi-classical approximations above, this equation simplifies further to

$$m\gamma(1 - \tau \partial_t^2 / \gamma)\partial_t \rho = k_B T(1 - \beta^2 \hbar^2 \partial_t^2 / 12) \partial_r^2 \rho \tag{14}$$

Interestingly, there is a temperature $T = \hbar(\gamma/3\tau)^{1/2}/2k_B$, at which Eq. (14) reduces to the classical diffusion equation with a diffusion constant $D = k_B T / m\gamma = \hbar(3\gamma\tau)^{1/2}/2m$. Surprisingly, their product $TD = \hbar c^2 / 8k_B \alpha$ is a universal constant, where $\alpha$ is the fine-structure constant.

It is interesting to compare the present results with the predictions of the linear theory for a free Brownian particle. In this case Eq. (1) reads $\dot{P} + \gamma P = F$, where $P \equiv m\dot{R}$ is the particle momentum. Using the Fourier transform, it is easy to derive the spectral density of the momentum fluctuations, $S_{PP} = S_{FF}/(\omega^2 + \gamma^2)$. Substituting here the spectral density of the Langevin force $S_{FF} = m\gamma I \hbar \omega \coth(\beta \hbar \omega / 2)$ and integrating over $\omega$ yields an expression for the momentum dispersion of the free classical Brownian particle

$$<P^2>_{eq} = \frac{1}{\pi}\int_0^\Omega \text{tr}(S_{PP})d\omega = \frac{3m\gamma}{\pi}\int_0^\Omega \frac{\hbar\omega\coth(\beta\hbar\omega/2)}{\omega^2+\gamma^2}d\omega \tag{15}$$

Note that an upper limit frequency $\Omega$ is introduced in Eq. (15), since the integral diverges logarithmically at high frequency. According to the Maxwell distribution (8), the equilibrium momentum dispersion on a classical particle is always $<p^2>_{eq} = 3mk_BT$. Substituting the latter in Eq. (15) provides after rearrangements an integral equation for the cut-off frequency $\Omega$

$$\beta\hbar\gamma\int_0^1 \frac{\coth(\beta\hbar\Omega\sqrt{x}/2)}{x+(\gamma/\Omega)^2}dx = 2\pi \tag{16}$$

Note that $<p^2>_{eq} = 3mk_BT$ is valid even for a free quantum Brownian particle. It is easy to estimate numerically from Eq. (16) that $\Omega \approx (2\pi\gamma k_BT/\hbar)^{1/2}$ is approximately equal to the geometric average of the friction $2\pi\gamma$ and lowest Matsubara $k_BT/\hbar$ frequencies. Obviously, $\Omega$ tends to infinity in the classical limit $\hbar \to 0$ or $T \to \infty$ but more interesting is that $\Omega$ vanishes at zero temperature or at vanishing friction coefficient. The latter is important for systems with small dissipation [8]. For instance, an estimate [9] for the quantum friction coefficient $m\gamma = 2\pi\hbar/\lambda^2$, where $\lambda$ is the mean-free path, leads to the conclusion that in this case the cut-off equals to the collision frequency, $\Omega/2\pi = (k_BT/m)^{1/2}/\lambda$.

Finally, it is posteriorly possible to enhance Eq. (12) to the case of full quantum Brownian motion. The effect of the quantum Brownian particle can be taken into account via the Bohm quantum potential $Q \equiv -\hbar^2\partial_r^2\rho^{1/2}/2m\rho^{1/2}$ [10]. Adding the latter to Eq. (12), the quantum-quantum Smoluchowski equation acquires the form

$$m\hat{\gamma}\partial_t\rho = \partial_r \cdot (\rho\partial_r U + \rho\partial_r Q + k_B\hat{T}\partial_r\rho) \tag{17}$$

This equation describes the Brownian motion of a quantum particle in quantum environment. It is nonlinear due to the quantum potential but it could be linearized to obtain

$$m\hat{\gamma}\partial_t\rho = \partial_r \cdot (\rho\partial_r U - \hbar^2 \partial_r^3 \rho / 4m + k_B \hat{T} \partial_r \rho) \tag{18}$$

Equation (18) shows that the quantum effect of the Brownian particle is similar to temperature. Indeed, one can introduce another quantum-quantum temperature operator $\hat{T} - \hbar^2 \partial_r^2 / 4mk_B$, which simplifies in the semi-classical limit to $T - \hbar^2(\partial_r^2 + m\partial_t^2 / 3k_B T)/4mk_B$. In the case of a free quantum Brownian particle, one can apply time and space Fourier transformations to Eq. (18) to get the following dispersion relation

$$(\hbar/2m)^2 q^4 + (\hbar\omega/2m)\coth(\beta\hbar\omega/2)q^2 - i\omega(\gamma + \tau\omega^2) = 0 \tag{19}$$

Using the solution of Eq. (19), $q^2 = (m\omega/\hbar)\{[\coth^2(\beta\hbar\omega/2) + 4i(\gamma/\omega + \tau\omega)]^{1/2} - \coth(\beta\hbar\omega/2)\}$, one can analyze dissipative spatial patterns, arising during the irreversible evolution in the so-called Madelung quantum fluid. For instance, the dispersion relation (19) tends at low frequency to the classical diffusive one $i\omega = q^2/m\gamma\beta$, while at high frequency without radiation ($\tau \equiv 0$) an interesting quantum dissipative structure is established with $q^2 = 2im\gamma/\hbar$.